\documentstyle[12pt,epsf]{article}

\if@twoside
 \else
 \fi

 \parskip \medskipamount
\begin{document}
\title{Thermalization algorithms \\
	for classical gauge theories}

\author{A. Krasnitz \\
	The Niels Bohr Institute \\
        Blegdamsvej 17, DK-2100 Copenhagen \\
        Denmark \\
	e-mail: krasnitz$@$hetws3.nbi.dk}
\date{July 1995}
\maketitle

\begin{abstract}
I propose a method, based on a set of Langevin equations, for bringing
classical
gauge theories to thermal equilibrium while respecting the set of
Gauss' constraints {\it exactly}. The algorithm is described in detail for
the SU(2) gauge theory with or without the Higgs doublet. As an example of
application, canonical average of the maximal Lyapunov exponent is
computed for the SU(2) Yang-Mills theory.
\end{abstract}
\vspace{-15cm}
\begin{flushright}
NBI-HE-95-25
\end{flushright}
\newpage

\section{Introduction}
In studying thermal properties of gauge theories it is useful to consider the
classical limit.
This is particularly true for classical lattice gauge theories whose
numerical simulation is increasingly used as a nonperturbative tool for
real-time dynamics of the gauge fields (sphaleron transitions,
properties of quark-gluon plasma
\cite{GRS,sphals1,sphals1a,sphals2,plasma,jaak}).

Simulation of a finite-temperature field theory should involve generating
the thermal ensemble of field configurations. In case of gauge theories,
configurations comprising the ensemble are subject to a set of local Gauss'
constraints, whose presence precludes straightforward application of standard
importance-sampling methods of ensemble generation in most cases. This work
deals with construction and implementation of constraint-respecting
thermalization algorithms for a number of physically important classical gauge
theories.

So far, three methods have been used to thermalize gauge theories.
One method consists of explicitly solving the constraints and
applying importance sampling to the remaining gauge-invariant
variables \cite{sphals2}. Its practical value is probably
limited to simple Abelian models.
For more complicated, non-Abelian theories the dynamics in a
gauge-invariant language is usually nonlocal and/or suffers from coordinate
singularities \cite{ginv}. The second method handles Gauss' constraints
approximately by adding to a Hamiltonian terms which penalize
deviations of the static charge density (SCD) from zero \cite{GRS,sphals1}.
Configurations with large SCD would then be unlikely to appear in a sample
generated by an importance sampling procedure. A special cooling procedure
is used to remove whatever SCD has nevertheless been generated in the system.
This method has its shortcomings too. Large SCD-suppressing
terms will dominate the energy functional and slow down the importance
sampling.
Besides, cooling may lead to deviations from the intended thermal ensemble.
Yet another method circumvents the need for preparing initial thermal
configurations by employing a self-consistent heat bath interacting with a
finite-size system at the boundaries in real time \cite{sphals2,rhb}. This
method, like the first
one, faces technical difficulties when applied to nonabelian theories or in
dimensions higher than one.

One would like to have a thermalization
algorithm formulated in terms of usual phase-space variables of a gauge
theory ({\it e.g.,} gauge potentials and color electric fields) and accurately
maintaining the Gauss' law. It turns out that such a method
can in several important cases be based on
Langevin equation with a specially designed multiplicative noise term.
That Langevin equations with multiplicative noise can be used for
thermalization of constrained systems has been known for quite a long time now.
This is how unitarity of link variables is maintained in Langevin simulations
of Euclidean gauge theories \cite{drum}. The unit-magnitude constraint on
fields in a sigma model can also be treated this way \cite{drum,habib}.
In what follows the same idea is employed in order to satisfy the set of
Gauss' constraints in a gauge theory. As we shall see, a natural choice of
algorithm in this case reflects the fact that Gauss' constraints are first
class in Dirac's terminology \cite{dirac}. For technical reasons,
it is convenient to use for our purpose second-order Langevin equations,
{\it i.e.,} of every pair of canonically conjugate phase-space
variables only one will be coupled to a heat bath by adding stochastic terms
to its equation of motion \cite{secord}.

The outlined strategy is discussed in some detail in the sections to follow.
In Section 2 I develop a Langevin formalism for Hamiltonian theories with
first-class constraints.
In Section 3 constraint-respecting coupling to a heat bath is constructed for
a number of gauge theories. Based on this construction, a Langevin dynamics
is formulated for the SU(2) lattice theory with and without a scalar doublet
in Section 4. Section 5 discusses numerical integration of the resulting system
of Langevin equations. Application of the Langevin algorithm is illustrated in
Section 6 where canonical average of the maximal Lyapunov exponent is computed
for the SU(2) Yang-Mills theory. Section 7 contains conclusions and outlook.
A brief outline of the ideas presented here was given in Ref.~\cite{prelim}.

\section{Langevin equations with first-class constraints}
In order to present the method in the most general form, I must briefly review
the basic phase-space properties of gauge theories. In doing so, I will
closely follow Ref.~\cite{ditsas}. Let a dynamical system be
described by $n$ coordinates $q_i$ and their respective conjugate momenta
$p_i$ forming a phase space $\Gamma$. Let these variables be subject to $m$
($m<n$) time-independent first-class constraints
$$C^\alpha(p,q)=0\approx\{H,C^\alpha\}\approx\{C^\beta,C^\alpha\},$$
where $H(p,q)$ is the Hamiltonian. The constraint surface ${\cal M}$ is a
$2n-m$-dimensional subspace of the phase space.
The weak equality sign, $\approx$, means,
as usual, "equal up to terms vanishing on the constraint surface".
A quantity $f(p,q)$ is physically meaningful (observable) if it is
gauge-invariant: $\{f,C^\alpha\}\approx 0$.
Obviously, the Hamiltonian $H$ is observable. Since any such $f$ is
constant along the gauge orbits
generated by $C^\alpha$ in ${\cal M}$, the physical subspace $\Gamma^*$ of
the phase
space is of dimension $2(n-m)$. Since the Poisson bracket of any two
observables
is again an observable, $\Gamma^*$ is by itself a phase space with a Poisson
bracket $\{\}^*$ defined for observables $u$ and $v$ as
$$\{u,v\}^*\equiv\{u,v\}|_{\cal M}.$$
It follows that there exists a canonical basis $p^*_i,q^*_i,1\leq i\leq m$
for functions in $\Gamma^*$ such that
$$\{p^*_i,p^*_j\}^*=\{q^*_i,q^*_j\}^*=0; \ \ \{p^*_i,q^*_j\}^*=\delta_{ij}.$$
All the observables (and $H$ in particular) on ${\cal M}$ depend exclusively on
$x^*\equiv p^*,q^*$.

Consider now Langevin dynamics in $\Gamma$. In its most general form the system
of Langevin equations is
\begin{equation}
\dot x_i=\{H,x_i\}+D_i(x)+g^j_i(x)\Gamma^j(t)), \label{mgen}\end{equation}
where I collectively denoted $p$ and $q$ by $x$. The system (\ref{mgen}) is
intentionally written as a generalization of Hamiltonian equations of motion:
the first term on the r.h.s. describes the canonical evolution, whereas the
other two terms arise from interaction of the dynamical system with a heat
bath. The $D_i$ terms are, in general, dissipative. The last term on the
r.h.s. is proportional to a white-noise random variable $\Gamma^j(t)$
with average zero and correlation
\begin{equation}
\langle\Gamma^j(t)\Gamma^k(t')\rangle=\delta_{jk}\delta(t-t').
\label{gacorr}\end{equation}
Here and in the following terms of this kind are to be understood in the
Stratonovich sense, unless otherwise stated. At the moment, I do not specify
the number of independent random variables $\Gamma^j(t)$.
Our goal of thermalizing the gauge
theory by means of (\ref{mgen}) will be achieved if {\it (a)} it preserves the
constraints, {\it i.e.,} generates a sequence of configurations on ${\cal M}$
from
an initial condition on ${\cal M}$, and {\it (b)} for a long evolution time
generates a sequence of configurations in $\Gamma^*$ distributed with canonical
probability density $\exp(-\beta H(p^*,q^*)$, where $\beta$ is the inverse
temperature. Since $\Gamma^j$ are random variables independent of $x$,
condition
{\it (a)} requires that
\begin{equation}
g^j_i\partial_iC^\alpha\approx 0.
\label{noviol}\end{equation}
Now I will show that condition {\it (b)} can be satisfied together with
{\it (a)} for a suitable choice of $D_i$ and $g^j_i$. In order to arrrive at
such an Ansatz, consider first the Fokker-Planck equation \cite{risken}
for the probability
density $W(x,t)$ in the full phase space $\Gamma$, which follows from
(\ref{mgen})
\begin{equation}
\left[\partial_t+\partial_i\left(\{H,x_i\}
+D_i+g^j_k\partial_kg^j_i-\partial_kg^j_kg^j_i\right)\right] W(x,t) =0
\label{fpe}\end{equation}
and require that it have $\exp(-\beta H(x))$ as its static solution.
To this end it is enough to choose $D_i=g^k_i G^k$, where
\begin{equation}
G^k=\partial_jg^k_j-\beta g^k_j\partial_jH.
\label{dij}\end{equation}
With this form of $D_i$ (\ref{mgen}) is constraint-preserving if (\ref{noviol})
holds. Our task is now reduced to finding a correct set of $g^k_j$ which in
turn would determine $D_i$. To further narrow down the search for a suitable
$g^k_j$ I require that (\ref{mgen}), averaged over realizations of random
variables, maps observables to observables. The equation of
motion for the average $\langle f(x,t)\rangle\equiv\int d^{2n}x W(x,t) f(x)$
of an observable $f(x)$ is obtained from (\ref{fpe}) integrated by parts over
the full phase space with boundary terms discarded:
\begin{equation}
\langle\dot
f\rangle=\langle\{H,f\}+\partial_k\left(g^j_ig^j_k\partial_if\right)
-\beta g^j_ig^j_k\partial_kH\partial_if\rangle.\label{avobs}\end{equation}
Gauge invariance of $\langle\dot f\rangle$ is ensured by choosing
\begin{equation}
g^j_i=M^{jk}(x)\{T^k(x),x_i\},\label{myg}\end{equation}
where $T^k$ is observable and the square matrix ${\cal M}M^{jk}$ is such that
$MM^T$
is also an observable quantity. Indeed, with this choice the quantity averaged
over on the r.h.s. of (\ref{avobs}) is
\begin{equation}
\{H,f\}+\{T^k,M^{jk}M^{jl}\{T^l,f\}\}-\beta M^{jk}M^{jl}\{T^k,H\}\{T^l,f\}.
\label{avrhs}\end{equation}
In addition, $g^j_i$ as given by (\ref{myg}) satisfies (\ref{noviol}). If the
initial configuration for (\ref{mgen}) lies on ${\cal M}$, (\ref{avrhs}), being
an
observable, depends exclusively on gauge-invariant canonical variables $x^*$.
Moreover, Poisson brackets $\{\}$ on $\Gamma$ can be replaced by their
counterparts $\{\}^*$ on $\Gamma^*$ everywhere in (\ref{avrhs}). Finally,
gauge invariance of (\ref{avrhs}) allows introduction of a gauge-invariant
probability density $W^*(x^*,t)$ on $\Gamma^*$ such that
$\langle f\rangle=\int d^{2(n-m)}x^*W^*f$ for any observable $f$ on ${\cal M}$.
In view
of (\ref{avobs}) and (\ref{avrhs}) $W^*$ satisfies the Fokker-Planck equation
\begin{equation}
\partial_tW^*=
\{W^*,H\}^*-\{T^k,M^{jk}M^{jl}(\{T^l,W^*\}^*+\beta W^*\{T^l,H\}^*)\}^*,
\label{fpegi}\end{equation}
written entirely in terms of $x^*$. It is obvious that (\ref{fpegi}) has
$\exp(-\beta H)$, with $H$ restricted to ${\cal M}$, as its static solution.
In the following I shall reserve the terms "generators" for $T^k$ and
"multiplier" for $M^{jk}$.

Substitution of (\ref{myg}) into (\ref{mgen}) gives the equation of motion for
an arbitrary variable $v$:
\begin{equation}
\dot v=\{H,v\}+\left(\{T^k,M^{jk}\}-\beta M^{jk}\{T^k,H\}+\Gamma^j\right)
M^{jl}\{T^l,v\}.
\label{dotv}\end{equation}
Note that, even though (\ref{mgen}) was written in a canonical basis,
(\ref{dotv}) is basis-independent. This property, familiar from the Hamiltonian
equations of motion, is useful whenever
the natural choice of independent dynamical variables is
not canonical, as is the case for Hamiltonian lattice gauge theories.
Another observation is that it is always possible to include in (\ref{dotv})
terms with $T^k=C^\alpha$. One one hand, this is allowed because the
constraints
are themselves first-class quantities. On the other hand, such an inclusion
obviously does not change the equations of motion for observables and is
therefore a matter of convenience.

There are important special cases of (\ref{myg}). If $M^{jk}=\delta^{jk}$,
the r.h.s. of (\ref{dotv}) is observable for any observable $v$ and separately
for any realization of $\Gamma^j$, not only on the average. In this case
the stochastic terms in (\ref{mgen},\ref{dotv}) may be thought of as arising
from interaction of the dynamical system with gauge-invariant degrees of
freedom
of the heat bath. Some of the examples to follow are of this kind.

The number $N$ of independent random variables $\Gamma^j$ and the functional
form of$T^k,M^{jk}$ must be specified so as to optimize the convergence of
$W^*(x^*,t)$ to $\exp(-\beta H)$.
If $N=2(n-m)$
it is always possible to choose $M^{jk}=\delta^{jk}$, $T^k=x^*_k$ for any
canonical basis $x^*$ in $\Gamma^*$. With this choice the heat-bath
fluctuations
described by $\Gamma^j$ generate the most general variation of variables in
$\Gamma^*$. The system (\ref{mgen}) then becomes a standard system of Langevin
equations with additive noise on $\Gamma^*$, known to guarantee convergence to
the equilibrium distribution. Thus the class of Langevin equations defined by
(\ref{dij}) and (\ref{myg}) is rich enough to achieve thermal equilibrium in
any
gauge theory. Convergent and numerically competitive algorithms are obtained
by taking $N=n-m$, $T^k=p^*_k$, {\it i.e.,} by coupling the heat
bath to a single variable in every canonically conjugate pair $p^*,q^*$
\cite{secord}.
In the following these algorithms are called second order.
In the next section I construct Langevin systems of this type for
a number of gauge field theories in the continuum. Later on I discuss
analogous algorithms for lattice gauge theories.

\section{Langevin systems in the continuum}

In this Section I study how the general prescription for finding a correct
heat bath coupling, as outlined in
Section 2, works in a number of gauge theories in the
continuum. It is convenient to set, with no loss of generality, $A_0=0$.
I will assume space dimension 3, which is the case for the most important
applications. It is implied that all phase-space variables carry space
coordinate labels, omitted here in order to simplify the notation.
It should be clear that Langevin systems of equations for a classical
field theory in the continuum are formal and cannot be put to immediate use:
the Rayleigh-Jeans divergency renders the classical thermal theory in the
continuum ill-defined. However, the continuum Langevin systems are useful as
a guidance for similar constructions on a lattice.

The simplest example is that of a free electromagnetic field. The Gauss'
law involves exclusively the electric fields $E_i$ and reads
$\partial_i E_i=0$. Hence an arbitrary variation of the gauge fields $A_i$
satisfies the Gauss' law. Such an arbitrary variation is generated by
$E_k,k=1,2,3$
and can be decomposed into
a variation of the physical, transversal part of the gauge potential and a
gauge
transformation. If $T^k=E_k$, $M^{jk}=\delta^{jk}$ in (\ref{dotv}),
it is straightforward to verify, by writing (\ref{dotv})
for $A$ and $E$ in momentum representation, that the resulting system of
equations is indeed second-order for the transversely polarized degrees of
freedom.

Next, I discuss the SU(2) Yang-Mills theory. The Gauss' constraints for color
gauge fields $A^\alpha_i$ and momenta $E^\alpha_i$ are
\begin{equation}
\partial_i E^\alpha_i-2\epsilon^{\alpha\beta
\gamma}A^\beta_iE^\gamma_i=0, \label{glawsu2}\end{equation}
where $\alpha,\beta,\gamma$ are adjoint color labels.
For fixed electric fields the most general variation of the gauge fields
can be written as
\begin{equation}
\delta A^\alpha_i=S_{ij}E^\alpha_j, \label{daym}\end{equation}
where $S_{ij}$ is a (space-dependent) gauge-invariant tensor.
Indeed, if the $3\times 3$ matrix $E^\alpha_j$ is invertible, as
is the case generically, we find
$S_{ij}=\delta A^\alpha_i (E^{-1})^\alpha_j$. For the Gauss' law
(\ref{glawsu2}) to hold, however, $S_{ij}$ must be symmetric.
Variations of $A$ of this form are generated by the six quantities
$T^{[ij]}=E^\alpha_i E^\alpha_j$, where
$[ij]$ labels ordered pairs with $1\leq i\leq j\leq 3$. The variables
$T^{[ij]}$ are gauge invariant, generically functionally independent, and
commuting with each other. Hence they can be chosen as momentum variables
in the observable subspace of the phase space. In fact, it is not difficult
to find the corresponding gauge-invariant conjugate variables. To this end
the original color electric fields are expressed in terms of $T^{[ij]}$ and
three Euler angles $\theta$ which parametrize an orthogonal matrix
$R^{\alpha a}(\theta)$ transforming $E^\alpha_i$ to some standard form
${\cal E}^a_i$. For instance, as pointed out by Goldstone and Jackiw
\cite{ginv}, it is possible to require that ${\cal E}^a_i$
be a product of an orthogonal matrix by a diagonal one.
Then for any $X=T^{[ij]},\theta_m$
$$P_{X_n}=A^\gamma_i{{\partial E^\gamma_i}\over{\partial X_n}}
-\epsilon^{\alpha\beta\gamma}{\partial\over{\partial X_n}}
\left(R^{\alpha a}(\theta)\partial_iR^{\beta a}(\theta)E^\gamma_i\right)$$
is the canonically conjugate variable.
Corresponding to the described canonical transformation is the generating
function
$$\Phi(A,X)=A^\gamma_iE^\gamma_i(X)-\epsilon^{\alpha\beta\gamma}
R^{\alpha a}(\theta)\partial_iR^{\beta a}(\theta)E^\gamma_i$$
whose partial derivatives give canonical conjugates of its arguments.
Gauge invariance of $P_T$ can be readily verified by performing a gauge
transformation
$$A^\alpha_i\rightarrow {\cal G}^{\alpha\beta}A^\beta_i-
\epsilon^{\alpha\beta\gamma}{\cal G}^{\gamma\sigma}
\partial_i{\cal G}^{\beta\sigma}; \ \ \
E^\alpha_i\rightarrow {\cal G}^{\alpha\beta}E^\beta_i$$
with an orthogonal ${\cal G}$. Consequently, the choice
of $T^{[ij]}=E^\alpha_i E^\alpha_j$ as generators and
$M^{[ij],[kl]}=\delta^{[ij],[kl]}$ as a multiplier
results in a second-order Langevin system in the physical subspace of the SU(2)
Yang-Mills theory. The system of equations for the standard variables can $A,E$
can be easily derived using (\ref{dotv}). It is obvious that the $E$ equations
are purely Hamiltonian.

Inclusion of scalar fields in the scheme presents no difficulty. The simplest
theory of this kind is scalar electrodynamics. In this case, the Gauss' law
is a relation between the electric field and the charge density of the
complex scalar field
\begin{equation}
C\equiv\partial_i E_i+i\left(\pi\phi-\phi^*\pi^*\right)=0,
\label{C}\end{equation}
where $\phi$ and $\pi$ are scalar field and momentum, respectively. The
constraint $C$ generates gauge transformations
$$A_i\rightarrow A_i-\partial_i\alpha; \ \ \phi\rightarrow\phi\exp(-i\alpha);
\ \ \pi\rightarrow\pi\exp(i\alpha),$$
where $\alpha$ is a gauge function. The most general variation of fields
consistent with (\ref{C}) consists of an arbitrary variation of $A_i$ and a
change in $\phi$ proportional to $\pi^*$:
\begin{equation}
\delta A_i = V_i; \ \ \delta\phi = S\pi^*. \label{deltas}\end{equation}
These are generated by $E_i,i=1,2,3$ and
$\Pi\equiv \pi^*\pi,$.
It is easy to identify the four (per space point) physical degrees of freedom
varied by (\ref{deltas}). Obviously, one can choose $E_i$ and the magnitude
$\Pi$ of $\pi$ as gauge-invariant momentum variables. The fifth momentum
variable, the phase $\theta$ of $\pi$, is clearly gauge-dependent.
Gauge-invariant canonical partners of $E_i$ and $\Pi$ follow from the
generating
function
$$\Phi(A,\phi,E,\Pi,\theta)=
E_iA_i+\pi(\Pi,\theta)\phi+\phi^*\pi^*(\Pi,\theta)-E_i\partial_i\theta.$$
These are
$$A_i-\partial_i\theta \ \ {\rm and}
\ \ {1\over{2\sqrt{\Pi}}}\left(\phi\exp(i\theta)+\phi^*\exp(-i\theta)\right),$$
respectively. Choosing $E_i,\Pi$ as generators and a unit multiplier
completes the construction of a second-order system for
the scalar electrodynamics.

My final example, the SU(2) theory with a scalar doublet,
describes a sector of the Standard Model. The Gauss' law
now reads
\begin{equation}
C^\alpha\equiv\partial_i E^\alpha_i-2\epsilon^{\alpha\beta
\gamma}A^\beta_iE^\gamma_i+i(\pi\sigma^\alpha\phi-\phi^*\sigma^\alpha\pi^*)=0.
\label{glhiggs}\end{equation}
Here the SU(2) spinors $\phi$ and $\pi$ are scalar fields and momenta,
$\sigma$ are Pauli matrices, and the rest of notation is obvious from
previous examples.

As before,
in order to find a suitable set of generators and multipliers for this system,
it is useful first to identify the physical degrees of freedom.
To this end, a point transformation of momenta is performed.
Introducing $\Pi\equiv\pi\pi^*$, the momentum transformation,
possible in the generic case $\Pi\neq 0$, reads
$$\pi=(\sqrt{\Pi} \ \ 0)U^\dagger(\theta); \ \ \ E_i^\alpha={1\over 2}{\rm Tr}
\left(\sigma^\alpha U(\theta){\cal E}_i^\beta\sigma^\beta U^\dagger(\theta)
\right)\equiv R^{\alpha\beta}(\theta){\cal E}_i^\beta,$$
where the Euler angles $\theta$ parametrize the unitary matrix $U$ rotating
$\pi$ to a standard form, here chosen to be $(\Pi^{1/2} \ \ 0)$. Matrix $R$,
defined by the last equality, is the orthogonal representation of $U$.
Only $\theta$ change under gauge
transformations, while $\Pi$ and the nine ${\cal E}$ variables are
gauge-invariant.
As in the previous examples, the canonical conjugates of $\Pi$ and ${\cal E}$
obtained with the help of a generating function
\begin{equation}
\Phi (A,\phi,\phi^*;{\cal E},\Pi,\theta)=A^\alpha_iE^\alpha_i({\cal E},\theta)
+\pi(\Pi,\theta)\phi+\phi^*\pi^*(\Pi,\theta)+{i\over 2}{\rm Tr}
\left({\cal E}^\alpha_i\sigma^\alpha U^\dagger(\theta)\partial_i
U(\theta)\right) \label{gfhiggs}\end{equation}
are again gauge invariant. These are
$$\phi_\Pi\equiv(\pi\phi+\phi^*\pi^*)/2\Pi$$
for $\Pi$ and
$${\cal A}^\alpha_i\equiv{1\over 2}{\rm Tr}\sigma^\alpha\left(U^\dagger(\theta)
A_i^\beta\sigma^\beta U(\theta)+iU^\dagger(\theta)\partial_iU(\theta)\right)$$
for ${\cal E}_i$. Thus ${\cal E}^\alpha_i,{\cal A}^\alpha_i$ and $\Pi,\phi_\Pi$
are the ten pairs of physical variables. It therefore makes sense to choose
${\cal E}$ and $\Pi$ as generators for the second-order Langevin system.
In this case, however, a convenient system of equations is obtained by using
a nontrivial multiplier $\Pi R^{\beta\gamma}$ for the generators
${\cal E}^\gamma_i$. With this choice terms proportional to the noise take
especially simple form
\begin{equation}
\Pi\Gamma^\gamma_j R^{\gamma\beta}\{{\cal E}^\beta_j,A^\alpha_i\}
=\Pi\Gamma^\alpha_i \label{aterm}\end{equation}
for the gauge potential. For the scalar field the corresponding term is
\begin{equation}
\Pi\Gamma^\gamma_j R^{\gamma\beta}\{{\cal E}^\beta_j,\phi\}
+\Gamma_\Pi\{\Pi,\phi\},\label{phiterm}\end{equation}
from which ${\cal E},R,\Pi$ can be readily eliminated in favor of the original
variables by noting that an arbitrary variation of the fields can be written as
\begin{equation}
\delta A^\alpha_i=\Pi\Gamma^\alpha_i, \ \ \
\delta\phi=(p+i\Sigma^\alpha\sigma^\alpha)\pi^*
\label{darb}\end{equation}
with real $\Gamma,p,\Sigma$, provided $|\pi|^2>0$ (generic case). Gauss' law
(\ref{glhiggs}) is only respected if
\begin{equation}
\Sigma^\alpha=-\epsilon^{\alpha\beta\gamma}
\Gamma^\beta_iE^\gamma_i. \label{sigmag}\end{equation}
Since variations (\ref{aterm},\ref{phiterm}) also respect the constraints,
they must be of the form (\ref{darb},\ref{sigmag}).
Comparing variation of $\phi_\Pi$ due to (\ref{darb}) with that due to
(\ref{phiterm}) yields $p=\Gamma_\Pi$. Finally, equality of terms linear in
$\Gamma^\alpha_i$ gives
$$\Pi R^{\beta\gamma}\{{\cal E}^\gamma_i,\phi\}
=-i\epsilon^{\alpha\beta\gamma}E^\gamma_i\sigma^\alpha\pi^*.$$

\section{Lattice Langevin equations: formulation}
The next step in our program is space discretization of constraint-preserving
thermalization scheme. Langevin equations of motion for lattice gauge theories
are constructed, following the analogy with the continuum case. Having in mind
applications to electroweak theory and to quark-gluon plasma, I will only write
down lattice equations for the two nonabelian theories already considered.
The reader should have no difficulty in writing a similar set of equations
for the abelian theories.

A convenient formalism for Hamiltonian ($A_0=0$) lattice gauge theory is that
of
Kogut and Susskind \cite{ks}. The configuration space consists of unitary
matrices $U_l$ forming the fundamental representation of the gauge group on
every link $l$ of a cubic lattice. In the following I shall use
the notation $j,{\hat{n}}$ for a link along a positive direction
${\hat{n}}$ originating at site $j$. Lattice analogs of electric fields are
link variables $E_l^{R\alpha}$ generating right covariant derivatives on the
group. Poisson brackets obeyed by $E_l^{R\alpha}$ with $U_l$ and among
themselves are only nonzero for variables residing on the same link, and then,
in the case of SU(2) gauge group
\begin{equation}
\{E_l^{R\alpha},U_l\}=-iU_l\sigma^\alpha; \ \
\{E_l^{R\alpha},E_l^{R\beta}\}=2\epsilon^{\alpha\beta\gamma}E_l^{R\gamma},
\label{pblat}\end{equation}
where $\sigma^\alpha$ are Pauli matrices.
The link variables $U_l$ and $E_l^{R\alpha}$ span the phase space.
Alternatively, one could replace $E_l^{R\alpha}$ as independent variables by
$E_l^{L\alpha},$ the generators of left covariant derivatives on the group.
On every link $l$ the transformation between the two sets of variables reads
\begin{equation}
E_l^{L\alpha}\sigma^\alpha=-E_l^{R\beta}U_l\sigma^\beta U^\dagger_l.
\label{rl}\end{equation}
As a matter of convention, I choose here $E_l^{R\alpha}$ as independent
variables. A useful property is
\begin{equation}
\{E_l^{R\alpha},E_l^{L\beta}\}=0.\label{useful}\end{equation}
Scalar doublet fields $\phi_j$, if added to the theory, are assigned,
together with their conjugate momenta $\pi_j$, to the sites $j$ of the lattice.
Gauge transformations of the variables sharing site $j$ are generated by
\begin{equation}
C^\alpha_j\equiv -\sum_{\hat{n}}
\left[E^{L\alpha}_{j, {\hat{n}}}
+E_{j-{\hat{n}},{\hat{n}}}^{R\alpha}\right]
- i\left(\pi_j\sigma^\alpha\phi_j-\phi_j^*\sigma^\alpha\pi_j^*\right)
\label{calpha}\end{equation}
(the scalar-field terms should be dropped from (\ref{calpha}) in case of pure
Yang-Mills theory). The set of Gauss' laws is $C^\alpha_j=0.$

Consider a Langevin system for the Yang-Mills theory first. Proceeding by
analogy with the continuum theory, I choose the gauge-invariant generators
$$T^{{[\hat{n}},{\hat{n'}]}}
=\sqrt{\gamma_E}E^{L\alpha}_{j, {\hat{n}}} E^{L\alpha}_{j, {\hat{n'}}}$$
with $1\leq {\hat{n}}\leq {\hat{n'}}\leq 3$, and a unit multiplier. The value
of
the friction coefficient $\gamma_E>0$ is arbitrary and can be tuned to optimize
the algorithm performance.
With the generators and the multiplier chosen, the Langevin
system is fully determined by the choice of a Hamiltonian.
The SU(2) Yang-Mills theory on the lattice is usually described
by a Kogut-Susskind Hamiltonian
\begin{equation}
H_{YM}={1\over 2}\sum_l E_l^{R\alpha} E_l^{R\alpha}
+ \sum_\Box\left(1-{1\over 2}{\rm Tr}U_\Box\right). \label{hym}\end{equation}
The second term in (\ref{hym}) is the standard plaquette term describing the
magnetic part of the energy. The Langevin system can now be written explicitly
using (\ref{dotv},\ref{pblat},\ref{rl},\ref{useful},\ref{hym}) (the site index
$j$ common to all variables is dropped to simplify the notation)
\begin{eqnarray}
\dot E_{\hat{n}}^{R\alpha}&=&-{i\over 2}{\rm Tr}
\left(\sigma^\alpha
U^\dagger_{\hat{n}}\sum_{\Box_{\hat{n}}}U_{\Box_{\hat{n}}}\right);\nonumber\\
\dot U_{\hat{n}}&=&-iE_{\hat{n}}^{R\alpha} U_{\hat{n}}\sigma^\alpha
-i\sqrt{\gamma_E}\sum_{\hat{n'}}
\left(G^{[{\hat{n}}{\hat{n'}}]}+\Gamma^{[{\hat{n}}{\hat{n'}}]}\right)
E^{L\beta}_{\hat{n'}} \sigma^\beta U_{\hat{n}},
\label{ymlang}\end{eqnarray}
Here $\Gamma^{\hat{n}},{\hat{n'}}$ is an independent random variable
corresponding to a given combination of a site $j$ and $[{\hat{n}}{\hat{n'}}]$,
\begin{equation}
G^{[{\hat{n}}<{\hat{n'}}]}\equiv -{{i\beta\sqrt{\gamma_E}}\over 2}
{\rm Tr}\left(E_{\hat{n'}}^{L\alpha}\sigma^\alpha
\sum_{\Box_{\hat{n}}}U_{\Box_{\hat{n}}}+n\leftrightarrow n'\right);
\ \ G^{nn}=i{{i\beta\sqrt{\gamma_E}}\over 2}
{\rm Tr}\left(E_{\hat{n}}^\alpha\sigma^\alpha
\sum_{\Box_{\hat{n}}} U_{\Box_{\hat{n}}}\right), \label{gym}\end{equation}
and $\Box_{\hat{n}}$ is any plaquette containing the link $U_{\hat{n}}$.
The $E^R$ equation of (\ref{ymlang}) has no stochastic or dissipative terms
because these variables commute with the generators.
Note that (\ref{ymlang}), beside satisfying the Gauss' law, automatically
preserves the unitarity of link matrices, as it should.

With the scalar field included, a Langevin system is again constructed by
analogy with the continuum case. Gauge invariant generators $\Pi\equiv |\pi|^2$
are introduced through
$$\Pi_j\equiv |\pi_j|^2; \ \ \pi_j\equiv\left(\sqrt{\Pi_j}\
0\right)V_j^\dagger;
\ \ {\cal E}_{j,\hat{n}}^\alpha\equiv
-{1\over 2}{\rm Tr}\left(\sigma^\alpha V_jU_{j,\hat{n}}E^{R\beta}_{j,\hat{n}}
\right)\equiv R^{\beta\alpha}_{j,\hat{n}} E^{R\beta}_{j,\hat{n}}.$$
The equations take a relatively simple form with multipliers
$\sqrt{\gamma}\Pi R^{\beta\alpha}_{j,\hat{n}}$ for
${\cal E}_{j,\hat{n}}^\alpha$ and $\sqrt{\gamma_\Pi}$ for
$\Pi$, where $\gamma$ and $\gamma_\Pi$ are arbitrary positive friction
coefficients.
With an addidion of the scalar field a representative Hamiltonian is
\begin{equation}
H_H=H_{YM}+\sum_j|\pi_j|^2+ \sum_{j,{\hat{n}}}|\phi_{j+{\hat{n}}}-U_{j,
{\hat{n}}}^\dagger\phi_j|^2+ \lambda\sum_jW\left(|\phi_j|^2\right),
\end{equation}
where $W$ is a local scalar field potential. In the resulting Langevin system
the $\pi$ and $E^R$ equations are simply the Hamiltonian ones, whereas for
$U$ and $\phi$ one obtains
\begin{eqnarray}
\dot U_{j,{\hat{n}}}&=&-i\left[E_{j,{\hat{n}}}^\alpha+\sqrt{\gamma}|\pi_j|^2
\left(\Gamma_{j,{\hat{n}}}^\alpha(t)+G_{j,{\hat{n}}}^\alpha\right)
\right] U_{j,{\hat{n}}}\sigma^\alpha;\nonumber\\
\dot\phi_j&=&\pi^*_j+\sqrt{\gamma_\Pi}\left(\Gamma^\Pi_j+G^\Pi_j\right)\pi^*_j+
i\sqrt{\gamma}\epsilon^{\delta\beta\rho}\sum_{\hat{n}}E^\beta_{j,
{\hat{n}}}\left(\Gamma^\rho_{j,{\hat{n}}}+G^\rho_{j,{\hat{n}}}\right)
U_{j,{\hat{n}}}\sigma^\delta U_{j,{\hat{n}}}^\dagger\pi^*_j,
\label{langh}\end{eqnarray}
where, as before, $\Gamma$ are mutually uncorrelated white noise variables, and
\begin{eqnarray}
G_{j,{\hat{n}}}^\alpha&=&-i\beta\sqrt{\gamma}
\left({{\partial H_H}\over{\partial U_{j,{\hat{n}}}}}
U_{j,{\hat{n}}}\sigma^\alpha+\epsilon^{\alpha\beta\gamma}
{{\partial H_H}\over{\partial \phi_j}}E^\beta_{j,{\hat{n}}}U_{j,{\hat{n}}}
\sigma^\gamma U_{j,{\hat{n}}}^\dagger\pi^*_j\right)+{\rm c.c};\nonumber\\
\ \ G^\Pi_j&=&-\beta\sqrt{\gamma_\Pi}{{\partial H_H}\over{\partial\phi_j}}
\pi^*_j +{\rm c.c} \label{gh}\end{eqnarray}
(summation over SU(2) spinor and adjoint indices only).

In the continuum theory, discussed in Section 3, the generators commuted among
themselves and therefore could be made, by a suitable canonical transformation,
canonical physical momenta. As a result, the continuum Langevin systems were
second-order systems in the physical subspace. Since the lattice generators
do not commute, the lattice Langevin system is not necessarily second-order in
this sense. Nevertheless, in the low-temperature regime ($\beta\gg 1$) the
continuum Langevin systems are recovered. Indeed, the lattice generators do
not commute due to the second relation of (\ref{pblat}). However, at low
temperature $E^R\sim \beta^{-1/2}$, as can be seen from (\ref{hym}), hence
the Poisson bracket of two $E^R$s is ${\cal O}(\beta^{-1/2})$,
much smaller than $\{p,q\}=1$ for any pair of canonically conjugate $p,q$.
Therefore, at low temperatures both (\ref{ymlang}) and (\ref{langh}) reduce
to second-order Langevin systems in the physical phase space.

\section{Lattice Langevin equations: numerical integration}
The final step in our construction of thermalization algorithms is finding
a suitable numerical integration scheme for a system of stochastic differential
equations
of the type (\ref{dotv}). It is best to use a scheme which respects the
Gauss' law. Such a scheme is indeed possible due to the following
useful property shared by all the Langevin systems constructed here.
Namely, with the conventional
choice of variables (space components of the gauge potential and the electric
fields) ithe Hamiltonian is a sum of a kinetic term $K$ commuting with all
the momentum variables $p$ (of which the lattice color electric fields are
a generalization), and a potential term ${\cal V}$ commuting with all the
coordinates $q$, with $K$ and ${\cal V}$ being {\it each} gauge-invariant.
Moreover, all the proposed generators $T^k$ of (\ref{dotv}) commute with all
the momenta. As a result, the Gauss' constraints $C^\alpha$ are conserved
{\it separately} by the momentum and coordinate equations of motion.
This property was used, although not explicitly stated, for purely canonical
equations in Ref.~\cite{sphals1}.
Moreover, since for the noise variables $\Gamma$ in (\ref{ymlang},\ref{langh})
all realizations in time are allowed, it is clear that this
property holds if $G$ terms in (\ref{ymlang},\ref{langh}) are arbitrary
functions of time, not necessarily
given by (\ref{gym}) and by (\ref{gh}). We conclude that the Gauss' law
holds {\it exactly}
for any combination of ({\it a}) integrating exactly the $E^R$ an $\pi$
equation of motion
while keeping $U$ and $\phi$ fixed, and ({\it b}) integrating exactly
the $U$ and $\phi$ equations while keeping $E^R$, $\pi$, and $G$ fixed
(the random variables
$\Gamma$ are also kept fixed for the duration of step ({\it b})). The freedom
in combining ({\it a}) and ({\it b}) can be used to optimize accuracy and
stability of the algorithm.

Some technical adjustments may have to be made in applying this general idea to
a specific gauge theory. Consider first the equation (\ref{ymlang})
for the SU(2) Yang-Mills theory. For arbitrary $\Gamma$ and $G$, the stochastic
terms on the r.h.s. involve in a nonlinear fashion all the link variables
for links emanating in positive directions from site $j$. Performing step
({\it b}) for such a system of coupled nonlinear equations is a very complex
task. If at any step we only allow a single off-diagonal
component of the symmetric tensor $\Gamma+G$ to be nonzero, a considerable
simplification will follow. Note that the general form of (\ref{ymlang}) is
(no implicit summation over $\hat{n}$ and $\hat{n'}$ in the following)
$$\dot U_{\hat{n}}=i\sum_{\hat{n'}}Z_{\hat{n}\hat{n'}}E^L_{\hat{n'}}
U_{\hat{n}},$$
where $Z_{\hat{n}\hat{n'}}$, symmetric under exchange of $\hat{n}$ with
$\hat{n'}$, is constant for the duration of step ({\it b}), and
$E^L{\hat{n}}\equiv E^{L\alpha}{\hat{n}}\sigma^\alpha$
obeys for fixed $E^R$ an equation of motion
$$\dot E^L{\hat{n}}=i\sum_{\hat{n'}}Z_{\hat{n}\hat{n'}}
\left[E^L{\hat{n}},E^L_{\hat{n'}}\right],$$
so that $\sum_{\hat{n}}\dot E^L{\hat{n}}=0$.
Consider now a special case where the $Z$ matrix has only one nonzero
off-diagonal element, {\it e.g.}, $Z_{\hat{n}\hat{n}}\equiv Z_{\hat{n}},$
$Z_{\hat{1}\hat{2}}\equiv Z,$
$Z_{\hat{1}\hat{3}}=Z_{\hat{2}\hat{3}}=0.$ Then $E^L_{\hat{3}}$, and,
equvalently, $E^L_{\hat{1}}+E^L_{\hat{2}}$
are time-independent, and the equations for $E^L$ and for $U$ are easy to
solve:
\begin{eqnarray}
U_{\hat{1},\hat{2}}(t)&=&\exp(-iZ(E^L_{\hat{1}}+E^L_{\hat{2}})t)
U_{\hat{1},\hat{2}}(0)
\exp(i(Z-Z_{\hat{1},\hat{2}})E_{\hat{1},\hat{2}}^{R\alpha}\sigma^\alpha t);
\nonumber\\
U_{\hat{3}}(t)&=&
U_{\hat{3}}(0)\exp(iZ_{\hat{3}}E_{\hat{3}}^{R\alpha}\sigma^\alpha t).\nonumber
\end{eqnarray}

In performing step ({\it b}) for the system (\ref{langh}) it helps to note that
the third term in the $\phi$ equation is completely fixed by requiring that it
compensates the violation of the Gauss' law generated by the $U$ equation. This
suggests the following procedure.
First the equations for the link matrices are integrated. Then violation
$\delta C^\alpha_j$ of the Gauss' law at site $j$, resulting from this change
of link matrices alone, is determined. Finally, the Gauss' law is restored by
the change in the scalar field:
\begin{equation}
\phi(t)-\phi_j(0)=\left[1+\sqrt{\gamma_\Pi}(\Gamma^\Pi_j+G^\Pi_j)
+{i\over{2|\pi_j|^2}}
\epsilon_{\alpha\beta\gamma}\delta C^\alpha_j
\sigma^\alpha\right]\pi^*_j.
\label{upstoch}\end{equation}
There is one caveat in this procedure: if $|\pi_j|^2$ is close to
zero, divisions by $|\pi_j|^2$ required for the scalar field updates in
(\ref{upstoch}) can be numerically unsafe. At the same time, comparison of
(\ref{upstoch}) with the $\phi$ equation of (\ref{langh}) shows that
$\delta C^\alpha_j/|\pi_j|^2$ approaches a finite value as
$|\pi_j|^2\rightarrow 0$. The remedy is therefore
to set a minimal value $\epsilon$ of $|\pi_j|^2$, and, whenever
$|\pi_j|^2<\epsilon$, replace $\delta C^\alpha_j/|\pi_j|^2$ by an estimate
obtained in the following way. First $|\pi_j|^2$ explicitly appearing in
the $U_{j,\hat n}$ equations of (\ref{langh}) is replaced by $\epsilon$.
Next, the $U_{j,\hat n}$ equations are integrated in order to find
$\delta C^\alpha_j$. The estimate is then $\delta C^\alpha_j/\epsilon$.
It is not difficult to show that, as a result of this replacement,
the Gauss' law is maintained approximately, its violation being at most
${\cal{O}}(\epsilon^2)$, which can be made completely negligible by an
appropriate choice of $\epsilon$.

Steps ({\it a}) and ({\it b}) must be combined in a suitable fashion to form
an accurate and stable integration algorithm. A simple scheme accurate to the
order 3/2 in the time step $\Delta$ is as follows. Initially $(t=0)$ a set of
random variables $\Gamma$ is generated. These remain constant for the duration
of the step. The integration step is schematically represented as
\begin{enumerate}
\item $E^R(0), \pi(0)\rightarrow E^R\left({\Delta\over 2}\right),
\pi\left({\Delta\over 2}\right)$ for fixed $U(0),\phi(0)$;
\item $U(0),\phi(0)\rightarrow U\left({\Delta\over 2}\right),
\phi\left({\Delta\over 2}\right)$  for fixed
$E^R\left({\Delta\over 2}\right),\pi\left({\Delta\over 2}\right),G(0)$;
\item $U(0),\phi(0)\rightarrow U(\Delta),\phi(\Delta)$ for fixed
$E^R\left({\Delta\over 2}\right),\pi\left({\Delta\over 2}\right),
G\left({\Delta\over 2}\right)$;
\item $E^R\left({\Delta\over 2}\right),\pi\left({\Delta\over 2}\right)
\rightarrow E^R(\Delta),\pi(\Delta)$ for fixed $U(\Delta),\phi(\Delta)$.
\end{enumerate}
Here the $\rightarrow$ symbol denotes exact integration. All the color and
space
indices have been dropped in order to simplify the notation. For a pure
Yang-Mills theory the $\phi$ and $\pi$ variables should be omitted. Note that
this scheme is a generalization of the simplest leapfrog algorithm to
which it reduces in the absence of stochastic terms (item 2 above then becomes
obsolete). If not for the $\Gamma$ terms present in the stochastic equations
the algorithm would have an ${\cal{O}}(\Delta^3)$ error
per step, similar to the simplest leapfrog. However, the noise terms $\Gamma$
must be implemented as random variables having average 0 and variance
$2/\Delta$ in order to approximate (\ref{gacorr}). The error per step is
therefore ${\cal{O}}(\Delta^{3/2})$.

Numerical tests, conducted at high ($\beta=2$) and low ($\beta=12$)
temperatures,
confirm that the algorithms in question indeed achieve the
correct thermalization while maintaining the Gauss' law with high accuracy.
The tests were performed on $12^3$ lattices with algorithm parameters
$\gamma_E=0.05$, $\Delta=0.01$ for the Yang-Mills theory and
$\gamma=0.04$, $\gamma_\Pi=0.2$, $\Delta=0.005$ in presence of the scalar
field. The scalar potential was
$\lambda(|\phi|^2-v^2)^2$, {\it i.e.,} $H_H$ corresponded to the bosonic
SU(2) sector of the Standard Model. The algorithm was tested with
$\lambda=0.5$, setting unit Higgs
to vector boson mass ratio, and $v^2=0.05$, corresponding to the
Higgs inverse mass of about $4.5$ lattice spacings.

\begin{figure}
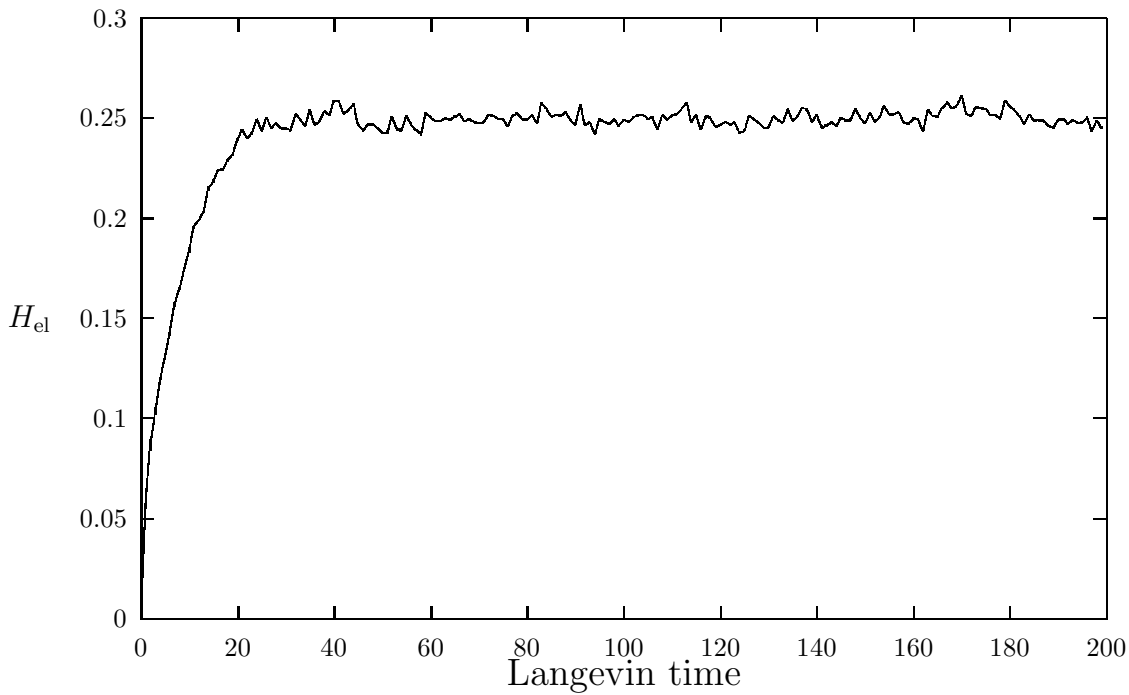

\setlength{\unitlength}{0.240900pt}
\ifx\plotpoint\undefined\newsavebox{\plotpoint}\fi
\sbox{\plotpoint}{\rule[-0.200pt]{0.400pt}{0.400pt}}%
%
\caption{Langevin time history of the color electric energy per site for the
SU(2) Yang-Mills theory on a $12^3$ lattice. The algorithm time step was 0.01.
Initial configuration had energy small compared to the average one for the
inverse temperature $\beta=12.$}
\label{evst}
\end{figure}

Figure~\ref{evst} illustartes the thermalization process for the SU(2)
Yang-Mills theory
described by (\ref{hym}). The initial configuration had zero electric fields
and
the energy far below the average one for the assigned inverse temperature
$\beta=12$. The system then rapidly reached thermal equilibrium.
An extremely small static
color charge per site generated by the algorithm is entirely
due to a limited computer accuracy and does not depend on the time step.
On a Cray-C90 processor with single-precision arithmetic the magnitude of the
spurious static charge was less than $4\times 10^{-12}$ per site
at the end of evolution shown in Figure~\ref{evst}.
Similar negligible amounts of constraint violation were observed in tests
of the Langevin system (\ref{langh}) with the scalar field.

Several criteria were used to judge how close the algorithms come to generating
canonical distributions for the two theories in question. In one instance,
the thermal average of an observable is known analytically. Namely, the
radial component of the scalar field momentum, $2{\rm Re}\pi\phi/|\phi|$
appears in $H_H$ quadratically and can be chosen as an independent physical
canonical variable. Hence the average radial kinetic energy per lattice site
should be $1/2\beta$. This exact result was indeed found within margins of the
measurement error in both low and high temperature regimes, with a $0.2\%$
accuracy.

Unfortunately, I am not aware of other observables with exactly known thermal
averages. For some quantities, however, perturbative estimates should be
reliable at low temperatures, opening another possibility of comparison
with numerical experiment. In particular, since the low-temperature system
is only weakly nonlinear, the average energy
per degree of freedom should be close to the equipartition value $1/\beta$.
At $\beta=12$ the measured average energy per degree of freedom is
$1.0091(7)/\beta$ for the Yang-Mills theory and $1.0053(7)/\beta$ with the
scalar fields.

Finally, the equilibrium statistical weight, $\exp(-\beta H)$, being a
function of the energy only, is conserved by the Hamiltonian evolution.
This property is used for the following consistency test of the algorithms.
An initial thermal configuration is subject to the Hamiltonian evolution,
and observables are measured at the beginning and at the end of the Hamiltonian
trajectory. For any observable, the two sets of measurements should yield the
same average if the generated distribution is indeed canonical. This test was
conducted for a number of observables of the Yang-Mills theory, namely, the
color electric energy
density, the topological charge density, and spatial Wilson loops of various
sizes. With the scalar field included, the radial kinetic energy density,
the Bricmont-Fr\"ohlich correlation function \cite{bfcorr} at various
distances,
and the "string bit"
$\phi^*_j U_{j,\hat{n}}\phi_{j+\hat{n}}/(|\phi^*_j||\phi_{j+\hat{n}}|)^{1/2}$
were added to this list. All the measurements show that the generated
distributions are close to the canonical ones. As an example, Table \ref{wtab}
summarizes the results for the spatial Wilson loops.
\begin{table}
\centerline{
\begin{tabular}{|cccll|} \hline
{system} & {size} & {$\beta$} & {initial} & {final} \\ \hline
{$H_{YM}$} & {$2\times 2$} & {12} & {0.7519(4)} & {0.7521(4)} \\
{$H_{YM}$} & {$3\times 3$} & {12} & {0.584(1)} & {0.584(1)} \\
{$H_{YM}$} & {$4\times 4$} & {12} & {0.436(2)} & {0.436(2)} \\
{$H_{YM}$} & {$5\times 5$} & {12} & {0.316(2)} & {0.318(2)} \\
{$H_{YM}$} & {$1\times 2$} & {2} & {0.1046(5)} & {0.1049(5)} \\
{$H_{YM}$} & {$2\times 2$} & {2} & {0.0450(8)} & {0.0450(7)} \\
{$H_{YM}$} & {$2\times 3$} & {2} & {0.0046(5)} & {0.0051(5)} \\
{$H_{H}$} & {$2\times 2$} & {12} & {0.7501(4)} & {0.7507(5)} \\
{$H_{H}$} & {$3\times 3$} & {12} & {0.580(1)} & {0.581(1)} \\
{$H_{H}$} & {$4\times 4$} & {12} & {0.432(2)} & {0.433(2)} \\
{$H_{H}$} & {$5\times 5$} & {12} & {0.312(2)} & {0.312(2)} \\
{$H_{H}$} & {$1\times 2$} & {2} & {0.1052(5)} & {0.1046(5)} \\
{$H_{H}$} & {$2\times 2$} & {2} & {0.0448(8)} & {0.0452(7)} \\
{$H_{H}$} & {$2\times 3$} & {2} & {0.0041(5)} & {0.0047(5)} \\
\hline
\end{tabular}}
\caption{Summary of measured canonical averages for spatial Wilson loops
of various sizes at the beginning (initial) and at the end (final) of
a Hamiltonian trajectory 50 time units long.}
\label{wtab}
\end{table}

\section{Application: maximal Lyapunov exponent of the SU(2) Yang-Mills theory}
As pointed out in the Introduction, lattice theories in the
classical approximation are especially valuable in dealing with dynamical
properties at finite temperature since in this case the arsenal of
nonperturbative tools applicable to a full quantum theory is extremely scarce.
Lyapunov exponents of a gauge theory belong to this category. According to
recent numerical work \cite{plasma}, the maximal Lyapunov exponents
$\lambda_{\rm max}$
of the classical SU(2) and SU(3) lattice gauge theories are approximately
independent of the lattice spacing and proportional to the average energy per
degree of freedom in a range of values of the latter. This approximate scaling
property of $\lambda_{\rm max}$ was suggested based on a set of initial
configurations
with zero electric energy and the values of link matrices randomly chosen from
a
closed vicinity of identity. Here I use the thermalization algorithm for the
SU(2) Yang-Mills theory described by (\ref{hym}) to compute
$\langle\lambda_{\rm max}\rangle,$
the {\it canonical ensemble} average of $\lambda_{\rm max}$ (notation
$\langle\rangle$ is used for canonical averages in the following). Taking this
average allows to directly establish a relation between the maximal Lyapunov
exponent and the temperature and to study deviations of
$\langle\lambda_{\rm max}\rangle$ from the scaling prediction of
Ref.~\cite{plasma} as compared to the statistical error of
$\langle\lambda_{\rm max}\rangle$.

\begin{figure}
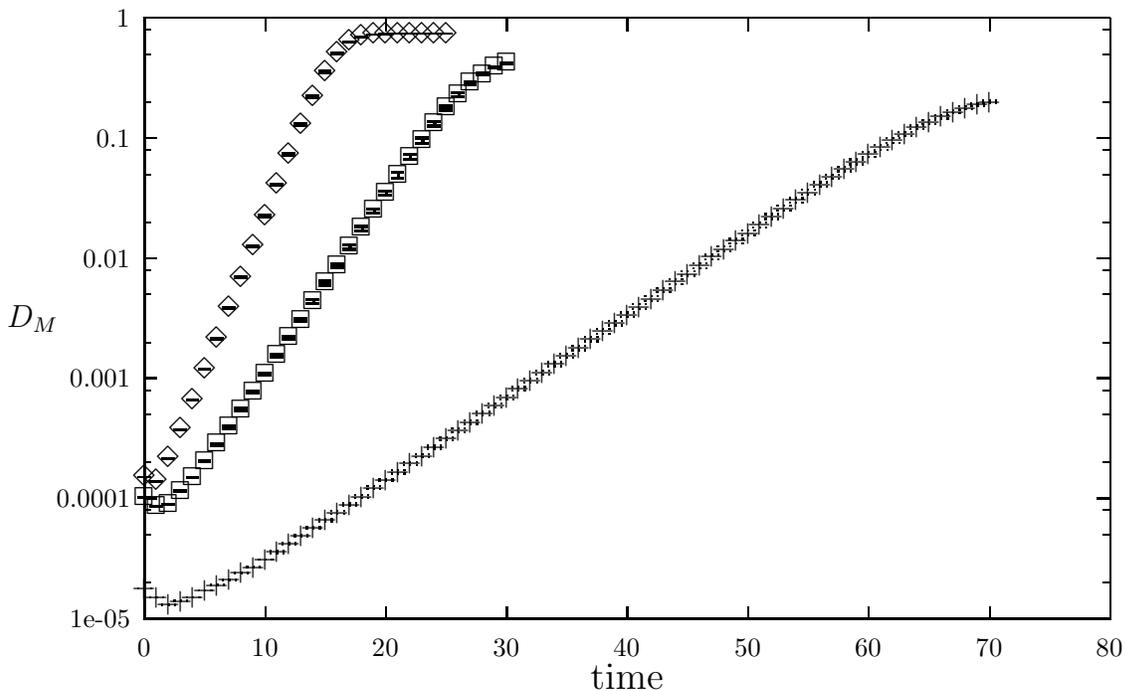

\setlength{\unitlength}{0.240900pt}
\ifx\plotpoint\undefined\newsavebox{\plotpoint}\fi
\sbox{\plotpoint}{\rule[-0.200pt]{0.400pt}{0.400pt}}%

\caption{Distance $D_M$ between two diverging trajectories as a function of
time for $\beta=2.5$ (diamonds), $\beta=4$ (squares), and $\beta=7.5$
(pluses) on a $12^3$ lattice. The error bars are smaller than the plotting
symbols}
\label{dmvst}
\end{figure}
The procedure used for measuring $\lambda_{\rm max}$ was as follows. A sample
of phase-space configurations corresponding to an inverse temperature $\beta$
was generated. Each member of the sample served as a reference initial
configuration. A neighboring configuration was then generated by applying
the algorithm briefly (in terms of Langevin time) to a reference one. Every
such
pair of nearby initial configurations was let evolve according to the
Hamiltonian equations of motion, and the distances between the two
configurations, defined as
$$D_E\equiv\sum_l |E_l^\alpha E_l^\alpha-{E'}_l^\alpha {E'}_l^\alpha|; \ \
D_M\equiv\sum_\Box|{\rm Tr}U_\Box-{\rm Tr}U'_\Box|,$$
\cite{plasma}, were monitored (the unprimed and primed variables correspond to
the reference and the neighboring configuration, respectively). A Langevin
trajectory between two consecutive Hamiltonian ones was long enough to
eliminate
autocorrelations of $\langle D_{E,M}(t)\rangle$ within the sample. The
measurements were performed for inverse temperatures $2\leq \beta\leq 10$.
A $20^3$ lattice was used for $\beta\geq 5$, and a $12^3$ lattice for higher
temperatures; this size reduction had no measurable effect on
$\langle\lambda_{\rm max}\rangle$ already at $\beta=4$. The Hamiltonian time
step was small enough to ensure conservation of energy to six significant
digits. For every value of the temperature 25 Hamiltonian trajectories were
performed.

\begin{figure}
\setlength{\unitlength}{0.240900pt}
\ifx\plotpoint\undefined\newsavebox{\plotpoint}\fi
\sbox{\plotpoint}{\rule[-0.200pt]{0.400pt}{0.400pt}}%
\begin{picture}(1800,1080)(0,0)
\font\gnuplot=cmr10 at 10pt
\gnuplot
\sbox{\plotpoint}{\rule[-0.200pt]{0.400pt}{0.400pt}}%
\put(220.0,113.0){\rule[-0.200pt]{365.204pt}{0.400pt}}
\put(220.0,113.0){\rule[-0.200pt]{0.400pt}{227.410pt}}
\put(220.0,113.0){\rule[-0.200pt]{4.818pt}{0.400pt}}
\put(198,113){\makebox(0,0)[r]{0}}
\put(1716.0,113.0){\rule[-0.200pt]{4.818pt}{0.400pt}}
\put(220.0,248.0){\rule[-0.200pt]{4.818pt}{0.400pt}}
\put(198,248){\makebox(0,0)[r]{0.1}}
\put(1716.0,248.0){\rule[-0.200pt]{4.818pt}{0.400pt}}
\put(220.0,383.0){\rule[-0.200pt]{4.818pt}{0.400pt}}
\put(198,383){\makebox(0,0)[r]{0.2}}
\put(1716.0,383.0){\rule[-0.200pt]{4.818pt}{0.400pt}}
\put(220.0,518.0){\rule[-0.200pt]{4.818pt}{0.400pt}}
\put(198,518){\makebox(0,0)[r]{0.3}}
\put(1716.0,518.0){\rule[-0.200pt]{4.818pt}{0.400pt}}
\put(220.0,652.0){\rule[-0.200pt]{4.818pt}{0.400pt}}
\put(198,652){\makebox(0,0)[r]{0.4}}
\put(1716.0,652.0){\rule[-0.200pt]{4.818pt}{0.400pt}}
\put(220.0,787.0){\rule[-0.200pt]{4.818pt}{0.400pt}}
\put(198,787){\makebox(0,0)[r]{0.5}}
\put(1716.0,787.0){\rule[-0.200pt]{4.818pt}{0.400pt}}
\put(220.0,922.0){\rule[-0.200pt]{4.818pt}{0.400pt}}
\put(198,922){\makebox(0,0)[r]{0.6}}
\put(1716.0,922.0){\rule[-0.200pt]{4.818pt}{0.400pt}}
\put(220.0,1057.0){\rule[-0.200pt]{4.818pt}{0.400pt}}
\put(198,1057){\makebox(0,0)[r]{0.7}}
\put(1716.0,1057.0){\rule[-0.200pt]{4.818pt}{0.400pt}}
\put(220.0,113.0){\rule[-0.200pt]{0.400pt}{4.818pt}}
\put(220,68){\makebox(0,0){0}}
\put(220.0,1037.0){\rule[-0.200pt]{0.400pt}{4.818pt}}
\put(437.0,113.0){\rule[-0.200pt]{0.400pt}{4.818pt}}
\put(437,68){\makebox(0,0){0.5}}
\put(437.0,1037.0){\rule[-0.200pt]{0.400pt}{4.818pt}}
\put(653.0,113.0){\rule[-0.200pt]{0.400pt}{4.818pt}}
\put(653,68){\makebox(0,0){1}}
\put(653.0,1037.0){\rule[-0.200pt]{0.400pt}{4.818pt}}
\put(870.0,113.0){\rule[-0.200pt]{0.400pt}{4.818pt}}
\put(870,68){\makebox(0,0){1.5}}
\put(870.0,1037.0){\rule[-0.200pt]{0.400pt}{4.818pt}}
\put(1086.0,113.0){\rule[-0.200pt]{0.400pt}{4.818pt}}
\put(1086,68){\makebox(0,0){2}}
\put(1086.0,1037.0){\rule[-0.200pt]{0.400pt}{4.818pt}}
\put(1303.0,113.0){\rule[-0.200pt]{0.400pt}{4.818pt}}
\put(1303,68){\makebox(0,0){2.5}}
\put(1303.0,1037.0){\rule[-0.200pt]{0.400pt}{4.818pt}}
\put(1519.0,113.0){\rule[-0.200pt]{0.400pt}{4.818pt}}
\put(1519,68){\makebox(0,0){3}}
\put(1519.0,1037.0){\rule[-0.200pt]{0.400pt}{4.818pt}}
\put(1736.0,113.0){\rule[-0.200pt]{0.400pt}{4.818pt}}
\put(1736,68){\makebox(0,0){3.5}}
\put(1736.0,1037.0){\rule[-0.200pt]{0.400pt}{4.818pt}}
\put(220.0,113.0){\rule[-0.200pt]{365.204pt}{0.400pt}}
\put(1736.0,113.0){\rule[-0.200pt]{0.400pt}{227.410pt}}
\put(220.0,1057.0){\rule[-0.200pt]{365.204pt}{0.400pt}}
\put(45,585){\makebox(0,0){$\lambda_{\rm max}$}}
\put(978,23){\makebox(0,0){$\langle H_{YM}\rangle/V$}}
\put(220.0,113.0){\rule[-0.200pt]{0.400pt}{227.410pt}}
\put(483,275){\raisebox{-.8pt}{\makebox(0,0){$\Diamond$}}}
\put(521,298){\raisebox{-.8pt}{\makebox(0,0){$\Diamond$}}}
\put(573,333){\raisebox{-.8pt}{\makebox(0,0){$\Diamond$}}}
\put(756,463){\raisebox{-.8pt}{\makebox(0,0){$\Diamond$}}}
\put(898,584){\raisebox{-.8pt}{\makebox(0,0){$\Diamond$}}}
\put(1139,783){\raisebox{-.8pt}{\makebox(0,0){$\Diamond$}}}
\put(1325,903){\raisebox{-.8pt}{\makebox(0,0){$\Diamond$}}}
\put(1576,1002){\raisebox{-.8pt}{\makebox(0,0){$\Diamond$}}}
\put(483.0,274.0){\rule[-0.200pt]{0.400pt}{0.723pt}}
\put(473.0,274.0){\rule[-0.200pt]{4.818pt}{0.400pt}}
\put(473.0,277.0){\rule[-0.200pt]{4.818pt}{0.400pt}}
\put(521.0,297.0){\rule[-0.200pt]{0.400pt}{0.723pt}}
\put(511.0,297.0){\rule[-0.200pt]{4.818pt}{0.400pt}}
\put(511.0,300.0){\rule[-0.200pt]{4.818pt}{0.400pt}}
\put(573.0,331.0){\rule[-0.200pt]{0.400pt}{0.964pt}}
\put(563.0,331.0){\rule[-0.200pt]{4.818pt}{0.400pt}}
\put(563.0,335.0){\rule[-0.200pt]{4.818pt}{0.400pt}}
\put(756.0,461.0){\rule[-0.200pt]{0.400pt}{1.204pt}}
\put(746.0,461.0){\rule[-0.200pt]{4.818pt}{0.400pt}}
\put(746.0,466.0){\rule[-0.200pt]{4.818pt}{0.400pt}}
\put(898.0,581.0){\rule[-0.200pt]{0.400pt}{1.686pt}}
\put(888.0,581.0){\rule[-0.200pt]{4.818pt}{0.400pt}}
\put(888.0,588.0){\rule[-0.200pt]{4.818pt}{0.400pt}}
\put(1139.0,778.0){\rule[-0.200pt]{0.400pt}{2.168pt}}
\put(1129.0,778.0){\rule[-0.200pt]{4.818pt}{0.400pt}}
\put(1129.0,787.0){\rule[-0.200pt]{4.818pt}{0.400pt}}
\put(1325.0,899.0){\rule[-0.200pt]{0.400pt}{1.686pt}}
\put(1315.0,899.0){\rule[-0.200pt]{4.818pt}{0.400pt}}
\put(1315.0,906.0){\rule[-0.200pt]{4.818pt}{0.400pt}}
\put(1576.0,999.0){\rule[-0.200pt]{0.400pt}{1.686pt}}
\put(1566.0,999.0){\rule[-0.200pt]{4.818pt}{0.400pt}}
\put(1566.0,1006.0){\rule[-0.200pt]{4.818pt}{0.400pt}}
\put(390,233){\makebox(0,0){$+$}}
\put(511,305){\makebox(0,0){$+$}}
\put(640,389){\makebox(0,0){$+$}}
\put(781,501){\makebox(0,0){$+$}}
\put(920,614){\makebox(0,0){$+$}}
\put(1046,744){\makebox(0,0){$+$}}
\put(1155,817){\makebox(0,0){$+$}}
\put(1249,893){\makebox(0,0){$+$}}
\put(1448,969){\makebox(0,0){$+$}}
\sbox{\plotpoint}{\rule[-0.400pt]{0.800pt}{0.800pt}}%
\put(390,245){\raisebox{-.8pt}{\makebox(0,0){$\Box$}}}
\put(511,310){\raisebox{-.8pt}{\makebox(0,0){$\Box$}}}
\put(640,382){\raisebox{-.8pt}{\makebox(0,0){$\Box$}}}
\put(781,506){\raisebox{-.8pt}{\makebox(0,0){$\Box$}}}
\put(920,618){\raisebox{-.8pt}{\makebox(0,0){$\Box$}}}
\put(1046,732){\raisebox{-.8pt}{\makebox(0,0){$\Box$}}}
\put(1155,815){\raisebox{-.8pt}{\makebox(0,0){$\Box$}}}
\put(1249,878){\raisebox{-.8pt}{\makebox(0,0){$\Box$}}}
\put(1448,979){\raisebox{-.8pt}{\makebox(0,0){$\Box$}}}
\sbox{\plotpoint}{\rule[-0.500pt]{1.000pt}{1.000pt}}%
\put(390,231){\usebox{\plotpoint}}
\put(390.00,231.00){\usebox{\plotpoint}}
\put(407.27,242.51){\usebox{\plotpoint}}
\put(424.13,254.60){\usebox{\plotpoint}}
\multiput(426,256)(17.270,11.513){0}{\usebox{\plotpoint}}
\put(441.33,266.22){\usebox{\plotpoint}}
\put(458.27,278.20){\usebox{\plotpoint}}
\multiput(462,281)(17.270,11.513){0}{\usebox{\plotpoint}}
\put(475.39,289.92){\usebox{\plotpoint}}
\put(492.66,301.44){\usebox{\plotpoint}}
\put(509.47,313.60){\usebox{\plotpoint}}
\multiput(510,314)(17.270,11.513){0}{\usebox{\plotpoint}}
\put(526.71,325.14){\usebox{\plotpoint}}
\put(543.60,337.20){\usebox{\plotpoint}}
\multiput(546,339)(17.270,11.513){0}{\usebox{\plotpoint}}
\put(560.77,348.85){\usebox{\plotpoint}}
\put(578.04,360.36){\usebox{\plotpoint}}
\multiput(582,363)(16.604,12.453){0}{\usebox{\plotpoint}}
\put(594.83,372.55){\usebox{\plotpoint}}
\put(612.10,384.07){\usebox{\plotpoint}}
\put(628.93,396.20){\usebox{\plotpoint}}
\multiput(630,397)(17.270,11.513){0}{\usebox{\plotpoint}}
\put(646.16,407.77){\usebox{\plotpoint}}
\put(663.43,419.29){\usebox{\plotpoint}}
\multiput(666,421)(16.604,12.453){0}{\usebox{\plotpoint}}
\put(680.22,431.48){\usebox{\plotpoint}}
\put(697.49,442.99){\usebox{\plotpoint}}
\multiput(702,446)(16.604,12.453){0}{\usebox{\plotpoint}}
\put(714.28,455.18){\usebox{\plotpoint}}
\put(731.55,466.70){\usebox{\plotpoint}}
\put(748.82,478.21){\usebox{\plotpoint}}
\multiput(750,479)(16.604,12.453){0}{\usebox{\plotpoint}}
\put(765.60,490.40){\usebox{\plotpoint}}
\put(782.87,501.92){\usebox{\plotpoint}}
\multiput(786,504)(16.604,12.453){0}{\usebox{\plotpoint}}
\put(799.66,514.11){\usebox{\plotpoint}}
\put(816.93,525.62){\usebox{\plotpoint}}
\multiput(822,529)(16.786,12.208){0}{\usebox{\plotpoint}}
\put(833.85,537.64){\usebox{\plotpoint}}
\put(850.67,549.78){\usebox{\plotpoint}}
\put(867.94,561.30){\usebox{\plotpoint}}
\multiput(869,562)(16.604,12.453){0}{\usebox{\plotpoint}}
\put(884.73,573.49){\usebox{\plotpoint}}
\put(902.00,585.00){\usebox{\plotpoint}}
\multiput(905,587)(17.270,11.513){0}{\usebox{\plotpoint}}
\put(919.18,596.64){\usebox{\plotpoint}}
\put(936.06,608.71){\usebox{\plotpoint}}
\multiput(941,612)(17.270,11.513){0}{\usebox{\plotpoint}}
\put(953.32,620.24){\usebox{\plotpoint}}
\put(970.12,632.41){\usebox{\plotpoint}}
\put(987.39,643.93){\usebox{\plotpoint}}
\multiput(989,645)(17.270,11.513){0}{\usebox{\plotpoint}}
\put(1004.52,655.64){\usebox{\plotpoint}}
\put(1021.45,667.63){\usebox{\plotpoint}}
\multiput(1025,670)(17.270,11.513){0}{\usebox{\plotpoint}}
\put(1038.65,679.24){\usebox{\plotpoint}}
\put(1055.51,691.34){\usebox{\plotpoint}}
\put(1072.78,702.85){\usebox{\plotpoint}}
\multiput(1073,703)(17.270,11.513){0}{\usebox{\plotpoint}}
\put(1089.85,714.64){\usebox{\plotpoint}}
\put(1106.83,726.56){\usebox{\plotpoint}}
\multiput(1109,728)(17.270,11.513){0}{\usebox{\plotpoint}}
\put(1123.98,738.24){\usebox{\plotpoint}}
\put(1140.89,750.26){\usebox{\plotpoint}}
\multiput(1145,753)(17.270,11.513){0}{\usebox{\plotpoint}}
\put(1158.16,761.77){\usebox{\plotpoint}}
\put(1175.18,773.64){\usebox{\plotpoint}}
\put(1192.22,785.48){\usebox{\plotpoint}}
\multiput(1193,786)(17.270,11.513){0}{\usebox{\plotpoint}}
\put(1209.32,797.24){\usebox{\plotpoint}}
\put(1226.28,809.19){\usebox{\plotpoint}}
\multiput(1229,811)(17.270,11.513){0}{\usebox{\plotpoint}}
\put(1243.55,820.70){\usebox{\plotpoint}}
\put(1260.52,832.64){\usebox{\plotpoint}}
\multiput(1265,836)(17.270,11.513){0}{\usebox{\plotpoint}}
\put(1277.61,844.40){\usebox{\plotpoint}}
\put(1294.47,856.47){\usebox{\plotpoint}}
\put(1311.32,868.55){\usebox{\plotpoint}}
\multiput(1312,869)(17.270,11.513){0}{\usebox{\plotpoint}}
\put(1328.59,880.06){\usebox{\plotpoint}}
\put(1345.48,892.11){\usebox{\plotpoint}}
\multiput(1348,894)(17.270,11.513){0}{\usebox{\plotpoint}}
\put(1362.65,903.77){\usebox{\plotpoint}}
\put(1379.61,915.71){\usebox{\plotpoint}}
\multiput(1384,919)(17.270,11.513){0}{\usebox{\plotpoint}}
\put(1396.71,927.47){\usebox{\plotpoint}}
\put(1413.98,938.98){\usebox{\plotpoint}}
\put(1430.81,951.11){\usebox{\plotpoint}}
\multiput(1432,952)(17.270,11.513){0}{\usebox{\plotpoint}}
\put(1448.04,962.69){\usebox{\plotpoint}}
\put(1464.95,974.71){\usebox{\plotpoint}}
\multiput(1468,977)(17.270,11.513){0}{\usebox{\plotpoint}}
\put(1482.09,986.40){\usebox{\plotpoint}}
\put(1499.36,997.91){\usebox{\plotpoint}}
\multiput(1504,1001)(16.604,12.453){0}{\usebox{\plotpoint}}
\put(1516.15,1010.10){\usebox{\plotpoint}}
\put(1533.42,1021.61){\usebox{\plotpoint}}
\put(1550.28,1033.71){\usebox{\plotpoint}}
\multiput(1552,1035)(17.270,11.513){0}{\usebox{\plotpoint}}
\put(1567.48,1045.32){\usebox{\plotpoint}}
\put(1576,1051){\usebox{\plotpoint}}
\end{picture}
\caption{Maximal Lyapunov exponent $\lambda_{\rm max}$ as a function of average
energy per site. Diamonds with the error bars smaller than plotting symbols are
canonical averages of the present work. Pluses and squares are
single-trajectory
results of Ref.~[6]. The dotted line is a linear fit through the origin.}
\label{lvse}
\end{figure}
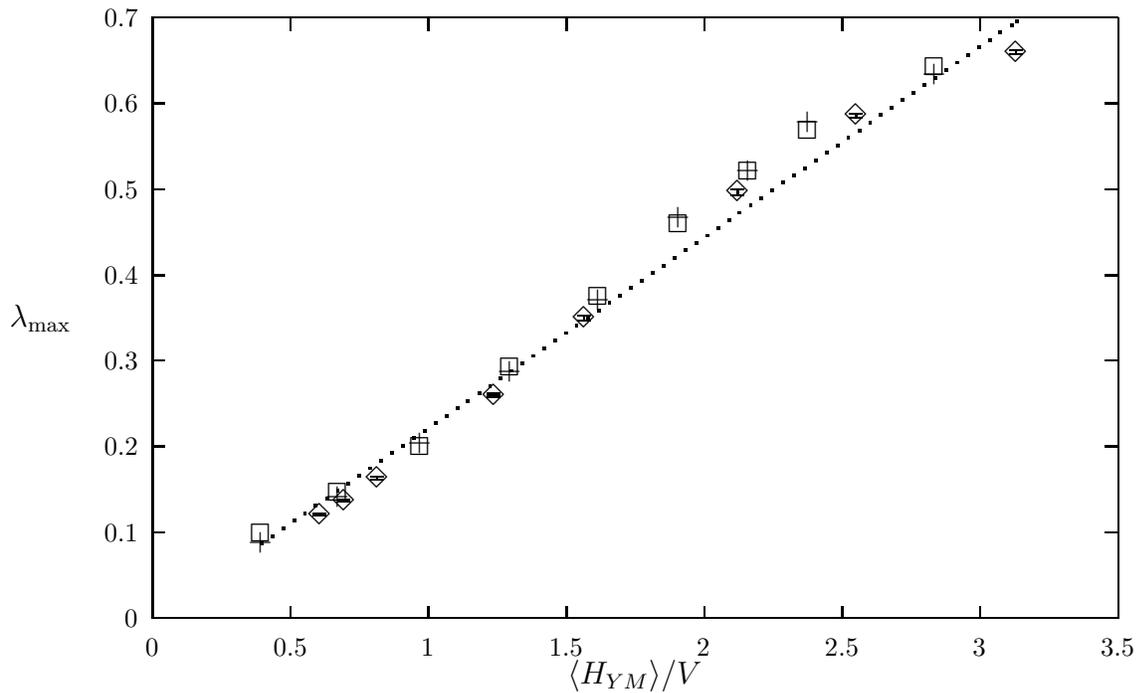
As functions of time, $\langle D_{E,M}(t)\rangle$ both
exhibit transient effects early on, then grow exponentially, and,
finally, saturate (Figure~\ref{dmvst}). The maximal Lyapunov exponent
$\lambda_{\rm max}$
is defined as the rate of the exponential growth of $D_{E,M}(t)$. A fit of
$D_{E,M}(t)$ to an exponential function of time yields
$\langle\lambda_{\rm max}\rangle$.  Within the measurement errors the values
of $\langle\lambda_{\rm max}\rangle$, determined from both definitions of
distance, coincide. Measurements in Ref.~\cite{plasma} suggested scaling of
$\langle\lambda_{\rm max}\rangle$ with the average
energy per site:
\begin{equation}
\langle\lambda_{\rm max}\rangle/\langle H_{YM}\rangle=\kappa.
\label{scaling}\end{equation}
\begin{table}
\centerline{\begin{tabular}{|llll|} \hline
{$\beta$} & {volume} & {$\langle H_{YM}\rangle$/volume}
& {$\langle\lambda_{\rm max}\rangle$} \\ \hline
{2} & {$12^3$} & {3.130} & {$0.659\pm 0.002$} \\
{2.5} & {$12^3$} & {2.551} & {$0.586\pm 0.002$} \\
{3} & {$12^3$} & {2.120} & {$0.497\pm 0.004$} \\
{4} & {$12^3$} & {1.566} & {$0.350\pm 0.003$} \\
{5} & {$20^3$} & {1.238} & {$0.260\pm 0.002$} \\
{7.5} & {$20^3$} & {0.8149} & {$0.163\pm 0.001$} \\
{8.75} & {$20^3$} & {0.6956} & {$0.137\pm 0.001$} \\
{10} & {$12^3$} & {0.6072} & {$0.115\pm 0.002$} \\
{10} & {$20^3$} & {0.6074} & {$0.120\pm 0.001$} \\
\hline
\end{tabular}}
\caption{Summary of measured canonical averages of $\lambda_{\rm max}$.}
\label{ltab}
\end{table}
According to Ref.~\cite{plasma} $\kappa\approx 2/9$ in the units of energy
and time used in this work. The best linear fit through the origin for the data
in Figure~\ref{lvse} is indeed very close to $2/9$. However, the goodness of
fit is very poor since, as Table~\ref{ltab} shows,
statistically significant deviations from (\ref{scaling}) are as large as 10\%.
Hence the estimate of $\kappa$ in the range of temperatures considered may not
be reliable. It is obvious from
dimensional considerations that any deviation from scaling
$\langle\lambda_{\rm max}\rangle\propto T$ implies dependence of
$\langle\lambda_{\rm max}\rangle$ on the lattice spacing. Lattice artifacts
tend to increase with the temperature due to the compact form of the lattice
magnetic term ({\it cf} (\ref{hym})). Thus, in order to safely establish the
value of $\kappa$, $\lambda_{\rm max}$ should be measured at temperatures
$T<0.1$. Larger lattice sizes may be required in order to control finite
size effects, clearly visible already at $T=0.1$. An accurate knowledge of
$\kappa$ through a low-temperature measurement is necessary in order to test
the relation of $\langle\lambda_{\rm max}\rangle$ to the plasmon damping rate
as suggested in Ref.~\cite{claim}.

\section{Concluding remarks}
Methods presented in this work enable one to accurately bring a variety of
classical lattice gauge theories to thermal equilibrium while respecting the
local charge conservation.
Due to their local nature, these methods easily lend themselves to vector
and parallel computer implementations\footnote{The codes are available upon
request from the author}. The algorithms can be applied to the
study of real-time properties of high-temperature gauge theories, where the
classical approach is by far the simplest if not the only one available.
There is a growing body of evidence that
some interesting  nonperturbative quantities can be reliably determined  by
classical real-time simulations and are not plagued by ultraviolet
divergencies inherent
in classical thermodynamics. The examples include approximate scaling of
thermalization
rate in (3+1)-dimensional Yang-Mills theories discussed in the previous
section,
as well as numerical evidence for the existence of continuum limit of
fermion-number violation rate in (1+1)-dimensional Abelian Higgs model
\cite{GRS,sphals2,sphals1a}. More recently, the SU(2) Yang-Mills thermalization
algorithm  described here was applied
to determine the baryon-number violation rate in the
Standard model at temperatures well above the electroweak phase transition
\cite{jaak}. The results show the existence of continuum limit for the rate
and its sensitivity to long-range properties of the theory
\footnote{A possibilty of this particular outcome of numerical experiment was
indicated in Ref.~\cite{bms}}.

It is possible to generalize the method to other gauge theories, most notably
to the physically important SU(3) Yang-Mills theory. Again, the algorithm
construction begins with finding a suitable set of gauge-invariant generators.
A candidate generator set is
$$T^{[ij]}\equiv E^\alpha_iE^\alpha_j; \ \ T^{[ijk]}\equiv
d^{\alpha\beta\gamma}
E^\alpha_iE^\beta_jE^\gamma_k,$$
where $E$ are the color electric fields, $d$ are the symmetric SU(3) constants,
and $i\leq j\leq k$. Gauge invariance of $T$ can be explicitly verified,
they commute with each other, and their number is 16, the number of the
physical
degrees of freedom. A lattice algorithm can be developed along these lines.

Finally, it should be noted that this work has concentrated on constructing
the algorithms and verifying their validity, rather than on tuning the
algorithm performance. The latter is postponed to subsequent work.
For the most obvious and important applications of the algorithms, however,
namely, real-time studies of gauge theories, generation of initial thermal
configurations typically consumes little time compared to canonical real-time
evolution. With this kind of application in mind, the algorithm optimization
can only lead to modest gains in computing time.

\section*{Acknowledgements}
I have had very beneficial discussions of this work with J.~Ambj{\o}rn,
P.~de~Forcrand, J.~Hetrick, A.~Kovner, L.~K\"arkk\"ainen,
Yu~Ming, N.~Nakazawa, H.B.~Nielsen, A.~Pasquinucci,
W.P.~Petersen, R.~Potting, S.E.~Rugh, J.~Smit, W.H.~Tang, and Y.~Watabiki. I am
grateful to B.~M\"uller for providing me with numerical data from
Ref.~\cite{plasma}.
This work was supported by the Danish Research Council under contract
No. 9500713. The bulk of the simulations was performed on the Cray-C92 and
the SP2 supercomputers at UNI-C.


\begin{thebibliography}{9}
\bibitem{prelim}A. Krasnitz, Nucl. Phys. B (Proc. Suppl.) 42 (1995) 885.
\bibitem{GRS}D.Yu.~Grigoriev, V.A.~Rubakov, and M.E.~Shaposhnikov,
Nucl. Phys. B326 (1989) 737.
\bibitem{sphals1}J. Ambj{\o}rn, T. Askgaard, H. Porter, and M.E. Shaposhnikov,
Nucl. Phys. B353 (1991) 346; Phys. Lett. B244 (1990) 479.
\bibitem{sphals1a}Ph. de Forcrand, A. Krasnitz, and R. Potting,
Phys. Rev. D50 (1994); \\
J. Smit and W.H. Tang, Nucl. Phys. B (Proc. Suppl.) 42 (1995) 590.
\bibitem{sphals2}A. Krasnitz and R. Potting, Phys. Lett. B318 (1993) 492.
\bibitem{plasma}T.S. Biro, C. Gong, B. M\"uller, and A. Trayanov, Int. J. Mod.
Phys. C5 (1994) 113; \\
B. M\"uller, and A. Trayanov, Phys. Rev. Lett. 68 (1992) 3387; \\
C. Gong, Phys. Rev. D49 (1994) 2642.
\bibitem{claim}T.S. Biro, C. Gong, and B. M\"uller, {\it Lyapunov exponents and
plasmon damping rate in nonabelian gauge theories,} Duke University report
DUKE-TH-94-73,  hep-ph/9409392.
\bibitem{jaak}J. Ambj{\o}rn and A. Krasnitz, {\it The classical sphaleron
transition rate exists and is equal to} $1.1(\alpha_wT)^4$, NBI report
NBI-HE-95-23, to be posted on hep-ph.
\bibitem{rhb}A. Krasnitz and R. Potting, Nucl. Phys. B410 (1993) 649.
\bibitem{drum}I.T. Drummond, S. Duane, and R.R. Horgan, Nucl. Phys. B220
(1983) 119; I.T. Drummond and R.R. Horgan,  Phys. Lett. B302 (1993) 271.
\bibitem{habib}S. Habib, Los Alamos preprint LA-UR-93-2105, gr-qc/9308022.
\bibitem{secord}A.M. Horowitz, Phys. Lett. B156 (1985) 89; Nucl. Phys. B280
(1987) 510; \\
S. Chaturvedi, A.K. Kapoor, and V. Srinivasan, Phys. Lett. B157 (1985) 400.
\bibitem{bms}D. B\"{o}deker, L. McLerran and A. Smilga,
{\it Really computing non-perturbative real time correlation functions},
University of Minnesota report TPI-MINN-95-08/T, hep-th/9504123.
\bibitem{dirac}P.A.M. Dirac, Proc. Roy. Soc. A249 (1958) 326.
\bibitem{ditsas}P. Ditsas, Ann. Phys. 167 (1986) 36.
\bibitem{ginv}J. Goldstone and R. Jackiw, Phys. Lett. B74, (1978) 81;
V. Baluni and B. Grossman, Phys. Lett. B78, (1978) 226; Yu.A. Simonov,
Sov. J. Nucl. Phys. 41 (1985) 835.
\bibitem{ks}J. Kogut and L. Susskind, Phys. Rev. D11 (1975) 395.
\bibitem{risken} H. Risken, {\it The Fokker-Planck equation: methods of
solution
and applications}, Springer, New York (1989).
\bibitem{bfcorr}J.~Bricmont and J.~Fr\"ohlich, Phys. Lett. B122 (1983) 73.
\end{thebibliography}
\end{document}